\documentclass[twocolumn,runningheads]{svjour2}
\smartqed  % flush right qed marks, e.g. at end of proof
\usepackage{graphicx}
\usepackage{mathptmx}      % use Times fonts if available on your TeX system
\journalname{Astrophysics and Space Science}
\begin{document}

\title{{\it INTEGRAL} serendipitous detection of the gamma-ray microquasar LS~5039
\thanks{Based on observations with {\it INTEGRAL}, an ESA project with
instruments and science data centre funded by ESA member states (especially
the PI countries: Denmark, France, Germany, Italy, Switzerland, Spain), Czech
Republic, and Poland, and with the participation of Russia and the USA.}
}



\author{P. Goldoni \and
        M. Rib\'o \and
        T. Di Salvo \and
        J.~M. Paredes \and
        V. Bosch-Ramon \and
        M. Rupen
}

\authorrunning{P. Goldoni, M. Rib\'o , T. Di Salvo , J.~M. Paredes, V. Bosch-Ramon, M. Rupen
} % if too long for running head

\institute{P. Goldoni \at
              APC/UMR 7164,
              Service d'Astrophysique - CEA
              Centre de Saclay, L'Orme des Merisiers
              91191, Gif-sur-Yvette Cedex, France\\
              Tel.: +33-(0)1-69085525\\
              Fax: +33-(0)1-69086577\\
              \email{pgoldoni@cea.fr}           %  \\
           \and
           M. Rib\'o, and J.~M. Paredes \at
              Departament d'Astronomia i Meteorologia,
              Universitat de Barcelona,
              Mart\'{\i} i Franqu\`es 1,
              08028, Barcelona, Spain\\
              Tel.: +34-934039225, +34-934021130\\
              Fax: +34-934021133\\
              \email{mribo@am.ub.es; jmparedes@ub.edu}  %  \\
           \and
           T. Di Salvo \at
              Dipartimento di Scienze Fisiche ed Astronomiche,
              Universit\`a di Palermo,
              via Archirafi 36,
              90123 Palermo, Italy\\
              Tel.: +39-091-6234251\\
              Fax: +39-091-6234281\\
              \email{disalvo@fisica.unipa.it}  %  \\
           \and
           V. Bosch-Ramon \at
              Max Planck Institut f\"ur Kernphysik,
              Saupfercheckweg 1,
              Heidelberg 69117, Germany\\
              Tel.: +49 6221516586\\
              Fax: +49 6221516324\\
              \email{vbosch@mpi-hd.mpg.de}  %  \\
           \and
           M. Rupen \at
              National Radio Astronomy Observatory,
              Array Operations Center,
              1003 Lopezville Road,
              Socorro, NM 87801, USA\\
              Tel.: +01 5058357248\\
              Fax: +01 5058357027\\
              \email{mrupen@aoc.nrao.edu}  %  \\
}

\date{Received: date / Accepted: date}

\maketitle

\begin{abstract}

LS~5039 is the only X-ray binary persistently detected at TeV energies by the
Cherenkov HESS telescope. It is moreover a $\gamma$-ray emitter in the GeV and
possibly MeV energy ranges. To understand important aspects of jet physics,
like the magnetic field content or particle acceleration, and emission
processes, such as synchrotron and inverse Compton (IC), a complete modeling
of the multiwavelength data is necessary. LS~5039 has been detected along
almost all the electromagnetic spectrum thanks to several radio, infrared,
optical and soft X-ray detections. However, hard X-ray detections above 20~keV
have been so far elusive and/or doubtful, partly due to source confusion for
the poor spatial resolution of hard X-ray instruments. We report here on deep
($\sim$300~ks) serendipitous {\it INTEGRAL} hard X-ray observations of
LS~5039, coupled with simultaneous VLA radio observations. We obtain a
20--40~keV flux of $1.1\pm0.3$~mCrab (5.9 ($\pm 1.6$) $\times$ 10$^{-12}$
erg cm$^{-2}$s$^{-1}$), a 40--100~keV upper limit of 1.5~mCrab (9.5 $\times$
10$^{-12}$ erg cm$^{-2}$s$^{-1}$), and typical radio flux densities of
$\sim$25~mJy at 5~GHz. These hard X-ray fluxes are significantly lower than
previous estimates
obtained with BATSE in the same energy range but, in the lower interval, agree
with extrapolation of previous {\it RXTE} measurements. The {\it INTEGRAL}
observations also hint to a break in the spectral behavior at hard X-rays. A
more sensitive characterization of the hard X-ray spectrum of LS~5039 from 20
to 100~keV could therefore constrain key aspects of the jet physics, like the
relativistic particle spectrum and the magnetic field strength. Future
multiwavelength observations would allow to establish whether such hard X-ray
synchrotron emission is produced by the same population of relativistic
electrons as those presumably producing TeV emission through IC.

%\PACS{First \and Second \and More}
\end{abstract}

\section{Introduction}
\label{intro}

Microquasars are radio-emitting X-ray binary sources, comprising a compact
object (black hole or neutron star) and a companion star, which produce jets
through the accretion-ejection mechanism and generate radiation along the
electromagnetic spectrum (e.g. \cite{mirabel99}). For several aspects, they
are scaled-down versions of Active Galactic Nuclei (AGN) or quasars, from
which their name comes. While less numerous than AGNs, microquasars allow for
a more detailed study of some of the AGN important properties thanks to both
their closer locations and their faster variability.

As X-ray binaries, microquasars generally produce electromagnetic radiation
mainly through accretion and emit most of their luminosity in the soft
(0.1--20~keV) and hard ($>$20\-~keV--1~MeV) X-ray domains. $\gamma$-ray
($E>1$~MeV) emission from microquasars is rarely observed. On the other hand,
a class of AGNs, blazars, are strong $\gamma$-ray emitters, presenting little
accretion disk radiation. The $\gamma$ radiation produced in blazars is
thought in general to come from IC scattering of synchrotron
photons by the high-energy particles of the jet, or Synchrotron Self Compton
(SSC). This emission, linked to the jet, is therefore particularly important
because it gives information on the poorly known ejection process and on the
jet composition and magnetic field. However, the study of the $\gamma$ radiation
is hampered by the limited knowledge available on the seed photon field that
is boosted by the high energy particles. Microquasars suffer to some extent
also the same problem. Moreover, the origin of hard X-ray emission in some
microquasars, like LS~5039, is still under debate since accretion disk
traces are not present and the jet could be the dominant component, like
in blazars. Nevertheless, there is no need at present of large Lorentz
factors to explain the $\gamma$-ray variability in microquasars, unlike
the case of their extragalactic relatives.

LS~5039 is a microquasar with slightly variable and extended radio emission
\cite{paredes00,paredes02}. It is a binary system with a $\sim$3.9 days
orbital period composed by an O6.5 V-type donor star and (likely) a black hole
as the compact object \cite{casares05}. LS~5039 is the only microquasar
detected at TeV energies by the HESS telescope up to now
\cite{aharonian05,aharonian06} and it is also a $\gamma$-ray emitter in the
GeV range \cite{paredes00,ribo02t} and possibly in the MeV range
\cite{strong01,collmar04}. The properties of the emission of LS~5039 are quite
unlike the ones of more common microquasars like for instance GRS~1915+105
where the thermal emission from the disk dominates the total luminosity.
Therefore, LS~5039 is a laboratory in which new aspects of the
accretion-ejection mechanism and radiative processes can be studied. The
origin of the emission has been proposed to be due either to an
accretion-powered relativistic jet (e.g. \cite{bosch05,dermer06,paredes06}) or
to the interaction between the relativistic wind of a young non-accreting
pulsar and the stellar wind of the donor star (\cite{dubus06} and references
therein), although the detection of a collimated jet from $\sim$1 to 100
milliarcsecond scales renders the second scenario unlikely.

\begin{figure*}
\centering
\includegraphics[width=0.8\textwidth]{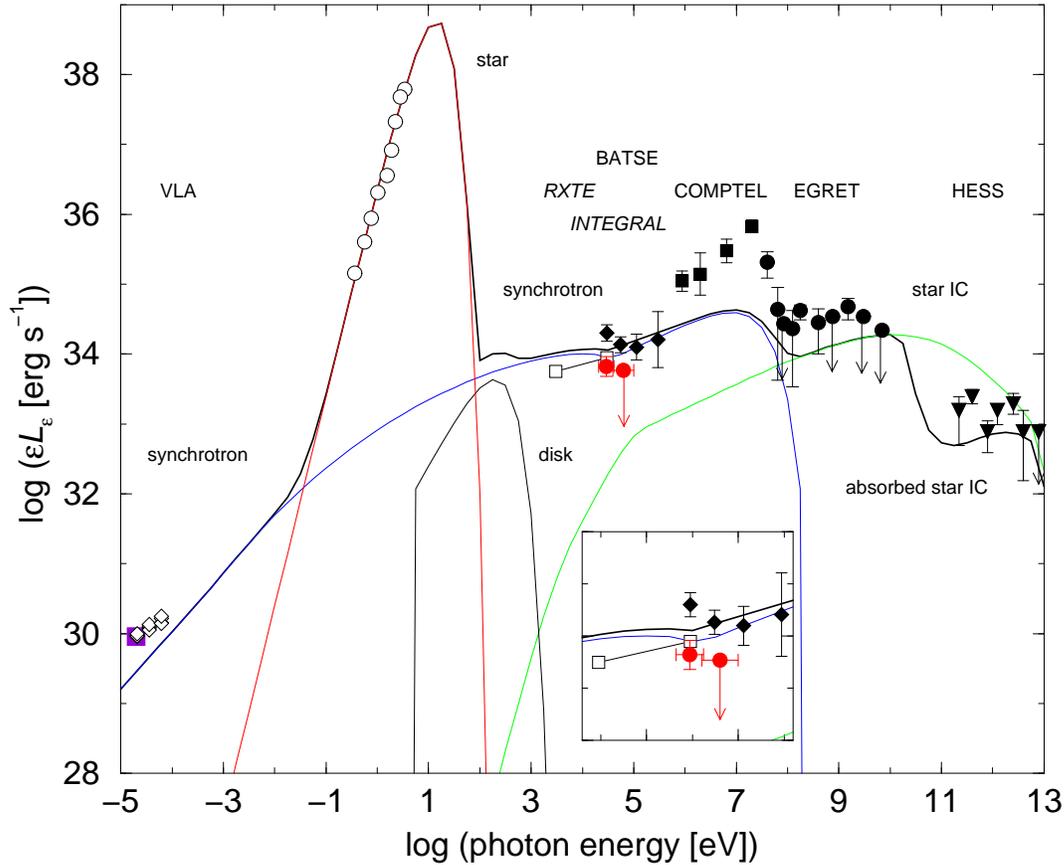}
\caption{Spectral energy distribution of LS~5039 from radio to TeV energies, a
model adapted from \cite{paredes06} is superimposed in which the most
important spectral components are shown. The references for the black
and white data points can also be found in \cite{paredes06}. We remark that
the MeV detection obtained with COMPTEL is very probably overestimated due
to source confusion. Our VLA detection is represented by a violet square,
while our {\it INTEGRAL} results are represented by red filled circles
(see zoom in the inset).}
\label{figsed} % Give a unique label
\end{figure*}

In the context of the accretion-powered model, the high energy emission is due
to IC scattering in the jet of a seed photon field. The seed photons could be
photons produced by synchrotron processes, which generate also the radio
emission at lower energies, or by the optical-UV photons of the donor star
(see Fig.~\ref{figsed}). The hard X-ray spectrum and the flux ratio between
hard X-rays and GeV-TeV emission differ depending on the dominant energy
losses, either synchrotron, Thomson IC or Klein-Nishina IC. Therefore, from
the hard X-ray spectral shape and comparing the hard X-rays to the GeV-TeV
flux, the magnetic field and even the dominant seed photon field (e.g. stellar
or synchrotron), can be inferred. A complete modeling of the emission of
LS~5039 is crucial for our understanding of the properties of the X-ray and
the $\gamma$-ray emitting region(s), as well as of the jet itself. Therefore,
hard X-ray emission is to be studied because it can shed light on the highest
energy part of the electron spectrum. In the following we will present
previous X-ray observations of LS~5039 and report serendipitous {\it INTEGRAL}
hard X-ray observations at the position of the source. This is presented
altogether with simultaneous VLA observations at 5~GHz, being also discussed
in the general context of the source.

\section{Previous soft and hard X-ray Observations}

In soft X-rays ($<$20~keV) LS~5039 has been detected several times (see
\cite{bosch05} and references therein). It is a weak source, $L_{\rm
X}\sim10^{34}$~erg~s$^{-1}$, with a power-law spectrum with photon index
$\Gamma\simeq1.5$. Its flux is variable (factor $\sim$2) with temporal scales
similar to the orbital period \cite{bosch05}. The variability seems to be
associated to the eccentric orbit of the system, $e\simeq0.35$
\cite{casares05}. The likely origin of the emission is synchrotron radiation
but a thermal origin from comptonization of a thermal plasma cannot at present
be excluded. In this last case X-ray and $\gamma$-ray emission would hardly
appear correlated.

However, the results of hard X-ray observations of LS~5039 have been so far
unclear, partly due to source confusion for the poor spatial resolution of
hard X-ray instruments. A BAT\-SE detection in the 20--100~keV range was
reported in the Earth Occultation catalog \cite{harmon04}, with an average
flux level of about 3.3~mCrab ($3.4\pm0.8$~mCrab at 20--40~keV and
$3.0\pm0.7$~mCrab at 40--70~keV\footnote{1 mCrab = 5.4 $\times$ 10$^{-12}$
erg cm$^{-2}$s$^{-1}$ (20--40~keV); 6.3 $\times$ 10$^{-12}$ erg cm$^{-2}$s$^{-1}$ (40--100~keV); 3.95 $\times$ 10$^{-12}$ erg cm$^{-2}$s$^{-1}$ (40--70~keV)}.
This is 2 to 3 times stronger than the
extrapolation of previously reported {\it RXTE}/PCA soft X-rays flux
measurements \cite{bosch05}. Moreover no detection has been reported in the
second ISGRI source catalog, with 5$\sigma$ upper limits of $\sim$1~mCrab in
the range 20--40~keV and $\sim$1.5~mCrab at 40--100~keV, for an exposure time
of 780~ks \cite{bird06}.

\section{Observations}

We report here on the results of serendipitous {\it INTEGRAL} observations
with contemporaneous radio monitoring of LS~5039. The hard X-ray observation
lasted 300~ks (i.e. $\sim$1 orbital period) and was performed in the standard
5$\times$5 dithering pattern. The target of the observation was the bright low
mass X-ray binary GX~17+2 at a 2.6 degree angular distance from LS~5039. The
source was thus in the Fully Coded Field of View of the IBIS/ISGRI hard X-ray
camera \cite{lebrun03} and therefore observed at full sensitivity in the
20--200~keV energy band. On the other hand, due to its smaller Field Of View,
the JEM-X soft X-ray (3--35~keV) instrument had only about 50\% sensitivity at
the position of LS~5039. We remark that the 12~arcmin angular resolution of
the IBIS/ISGRI camera allows to separate LS~5039 from nearby objects and that
the 3$\sigma$ detection sensitivity for a 300~ks ISGRI observation is
$\sim$1~mCrab. At the same time we obtained contemporaneous radio snapshots at
4.86~GHz with the NRAO\footnote{The National Radio Astronomy Observatory is a
facility of the National Science Foundation operated under cooperative
agreement by Associated Universities, Inc.} VLA in order to monitor the source
behavior. The observation logs are shown in Tables~\ref{INTEG} and \ref{VLA},
respectively

\begin{table}[t]
\caption{{\it INTEGRAL}/ISGRI observation log and results.}
\centering
\label{INTEG}
\begin{tabular}{lcc}
\hline\noalign{\smallskip}
Obs. Date              & 20--40~keV flux & 40--100~keV flux\\
(MJD)                  & (mCrab)         & (mCrab)         \\
\tableheadseprule\noalign{\smallskip}
53665.08--53669.36 & $1.1\pm0.3$     & $<1.5$ \\
\noalign{\smallskip}\hline
\end{tabular}
\end{table}

\begin{table}[t]
\caption{VLA observation log and results. Each measurement corresponds to a 5
minute snapshot observation.}
\centering
\label{VLA}
\begin{tabular}{lc}
\hline\noalign{\smallskip}
Obs. Date & 4.86~GHz flux density\\
(MJD)     & (mJy)           \\
\tableheadseprule\noalign{\smallskip}
53667.07  & $25.1\pm0.2$ \\
53667.86  & $25.1\pm0.1$ \\
53668.96  & $24.4\pm0.3$ \\
\noalign{\smallskip}\hline
\end{tabular}
\end{table}

\section{Results}

We analyzed the {\it INTEGRAL} observations using the standard analysis
pipeline \cite{goldwurm03} and we produced images in the 20--40 and
40--100~keV bands for the whole observation. No clear detection of LS~5039 was
found, in contrast with the expected $\sim$10$\sigma$ detection for a
$\sim$3~mCrab source according to \cite{harmon04}. After a careful subtraction
of all the bright sources in the field of view, we were able to find at the
position of LS~5039 a $\sim$4$\sigma$ excess in the 20--40~keV energy band. If
this excess is real, then the implied flux is $\sim1.1\pm0.3$~mCrab. No excess
is detected in the 40--100~keV band with a 3$\sigma$ upper limit of 1.5~mCrab.
We note that these values are in agreement with the upper limits derived from
the absence of the source in the most recent ISGRI source catalog
\cite{bird06}. Unfortunately, the weakness of the detection hampers the
possibility of searching for an orbital modulation at hard X-ray energies.

In soft X-rays, the off-axis position of the source in JEM-X and the presence
of the noise induced by GX~17+2 prevented the source detection. This result is
compatible with an extrapolation of the possible detection in the 20--40~keV
band. Our results are also broadly consistent with those presented by
\cite{bosch05} at soft X-ray energies.

The snapshot observations we obtained with the VLA showed the source at a
moderate level of activity with an average flux of about 25~mJy, in agreement
with historic values of 23--26~mJy at 5~GHz \cite{marti98}.

The observation results are quoted in Tables~\ref{INTEG} and \ref{VLA} and are
plotted in Fig.~\ref{figsed} (see the inset to evaluate the hard X-rays 
results).

\section{Discussion and conclusions}

Our results indicate that LS~5039 is a weak hard X-ray emitter, detected at
$\simeq$1~mCrab in the 20--40~keV energy range and with a flux below 1.5~mCrab
in the 40--100~keV interval. These results are in agreement with extrapolation
of {\it RXTE}/PCA data and with previous upper limits from the ISGRI source
catalog. However, they are $\sim$3 times lower than previously reported BATSE
values. The {\it INTEGRAL} measurements can be considered as an average along
an orbital cycle. Therefore, the mentioned discrepancy cannot be due to
orbital variability. Alternatively, it could be due to long-term variations in
the hard X-ray flux of LS~5039, as observed in soft X-rays and in the 
equivalent width of the H$\alpha$ line and thought to be caused by long-term
variations of the wind of the primary \cite{reig03}.

On the other hand, the radio flux density measurements indicate that the
source was not experiencing a particular low level of activity. Nevertheless,
we recall that radio emission variations are small and that no long-term
variability has yet been detected at radio wavelengths, suggesting that the
radio and X-ray emission are not correlated.

Although these results may allow to refine the jet models proposed by
\cite{paredes06}, see Fig.~\ref{figsed}, the modeling of the radiation in the
whole spectral range is not straightforward. Actually, the relatively hard
observed TeV spectrum \cite{aharonian06} and the X-ray fluxes slightly higher
than those observed at TeV energies do not fit in the one-zone leptonic
scenario (the particle spectrum at the energies of interest would be dominated
by synchrotron losses  instead of by IC Klein Nishina ones, rendering the TeV
spectrum much softer than observed). Therefore, the soft X-rays and the
TeV photons are probably coming from different regions. In fact, the detection
between 20--40~keV and the upper limit in the range 40--100~keV presented here
hint to a change in the spectral behavior of the source. This suggests that
hard X-rays could be due to synchrotron emission that is produced by the same
relativistic electrons that produce TeV radiation through IC. 

A longer pointed {\it INTEGRAL} or {\it Suzaku} observation is needed to
detect any eventual orbital hard X-ray variability, and to detect the source
in the 40--100~keV energy range. Only simultaneous soft/hard X-ray and TeV
observations would permit to obtain enough spectral and temporal information
to fully characterize the relativistic electron spectrum, the seed photon
field and the magnetic field strength, and to establish whether the X-ray and
the TeV emission have indeed their origin in the same population of electrons.

\begin{acknowledgements}
M.R. has been supported by the French Space Agency (CNES) and is  being
supported by a {\em Juan de la Cierva} fellowship from the Spanish Ministerio
de Educaci\'on y Ciencia (MEC).
V.B-R. thanks the Max-Planck-Institut f\"ur Kernphysik for its support and
kind hospitality.
M.R., J.M.P and V.B-R.acknowledge support by DGI of MEC under grant
AYA2004-07171-C02-01, as well as partial support by the European Regional
Development Fund (ERDF/FEDER).
\end{acknowledgements}

\end{document}